\documentclass[aps,prl,preprint,superscriptaddress,amssymb,footinbib,showpacs]{revtex4}

\usepackage{amsmath, bm, natbib}
\usepackage[english]{babel}
\usepackage{graphicx, subfigure, float}
\usepackage{colordvi}

\newcommand{\Eq}[1]{Eq. (\ref{#1})}
\newcommand{\Fig}[1]{Fig. \ref{#1}}
\newcommand{\pref}[1]{(\ref{#1})}
\newcommand{\quant}[2]{#1 \, \mathrm{#2}}

\renewcommand{\vec}[1]{\mbox{\boldmath$#1$}}

\begin{document}

\bibliographystyle{apsrev}

\title{Detecting recurrence domains of dynamical systems \\
    by symbolic dynamics}

\author{Peter beim Graben}
\email{peter.beim.graben@hu-berlin.de}
\affiliation{Dept. of German Language and Linguistics,
Humboldt-Universit\"at zu Berlin, Germany}
\affiliation{Bernstein Center for Computational Neuroscience Berlin, Germany}
\affiliation{Cortex Project, INRIA Nancy Grand Est, France}

\author{Axel Hutt}
\affiliation{Cortex Project, INRIA Nancy Grand Est, France}


\date{\today}

\begin{abstract}
We propose an algorithm for the detection of recurrence domains of complex dynamical systems from time series. Our approach exploits the characteristic checkerboard texture of recurrence domains exhibited in recurrence plots (RP). In phase space, RPs yield intersecting balls around sampling points that could be merged into cells of a phase space partition. We construct this partition by a rewriting grammar applied to the symbolic dynamics of time indices. A maximum entropy principle defines the optimal size of intersecting balls. The final application to high-dimensional brain signals yields an optimal symbolic recurrence plot revealing functional components of the signal.
\end{abstract}


\maketitle

States of complex dynamical systems often dwell for relatively long time in a particular domain of their phase spaces before the trajectory moves into another region. This is the case for metastability \cite{LarraldeLeyvraz05} and several kinds of instability such as saddles that are connected by heteroclinic trajectories \cite{HuttRiedel03, RabinovichEA08} or, e.g., the ``wings'' of the Lorenz attractor that are centered around its unstable foci \cite{Lorenz63}. According to Poincar\'{e}'s famous \emph{recurrence theorem} \cite{Poincare1890}, we could refer to such behavioral regimes as to \emph{recurrence domains} of a dynamical system. The detection of recurrence domains has become increasingly important in recent time in several applications such as spin glasses \cite{LarraldeLeyvraz05}, molecular configurations \cite{DeuflhardWeber05}, in the geosciences \cite{FroylandPadbergEA07} and in the neurosciences \cite{AllefeldAtmanspacherWackermann09, Friston97a, HuttRiedel03, RabinovichHuertaLaurent08}.

For the identification of recurrence domains from time series, their characteristic slow time scales have been separated from the fast dynamics of phase space trajectories by several clustering algorithms \cite{DeuflhardWeber05, AllefeldAtmanspacherWackermann09, LarraldeLeyvraz05, Froyland05, GaveauSchulman06, HuttRiedel03}. One method, sometimes called \emph{Perron clustering} \cite{DeuflhardWeber05}, starts with an \emph{ad hoc} partitioning of the system's phase space that leads to an approximate Markov chain description \cite{DeuflhardWeber05, AllefeldAtmanspacherWackermann09, LarraldeLeyvraz05, Froyland05, GaveauSchulman06}. Applying spectral clustering methods to the resulting transition matrix yields the time scales of the process, while their corresponding (left-)eigenvectors allow the unification of cells into a partition of
metastable states \cite{AllefeldAtmanspacherWackermann09, GaveauSchulman06}. Another approach by Hutt and Riedel \cite{HuttRiedel03} utilizes the slowing-down of the system's trajectory within saddle sets by means of phase space clustering.

Several of such methods are numerically rather time-consuming. For instance, Markov chain modeling requires an \emph{ad hoc} partitioning of the complete system's phase space into equally populated cells, from which transition probabilities must be estimated by counting measures. Subsequent spectral clustering methods perform various matrix multiplications and clustering techniques. All these algorithms are numerically rather expensive as illustrated in \cite{AllefeldAtmanspacherWackermann09}.

In this Letter, we propose a parsimonious algorithm for detecting recurrence domains from measured or simulated time series. Our starting point is Eckmann et al.'s \cite{EckmannOliffsonRuelle87} recurrence plot (RP) method for visualizing Poincar\'{e}'s recurrences. The proposed method is numerically less time-consuming and advantageous especially for high-dimensional data since it simply exploits the recurrence structure of the system's dynamics.

\begin{figure}[H]
\centering
\subfigure[]{\includegraphics[width=60mm]{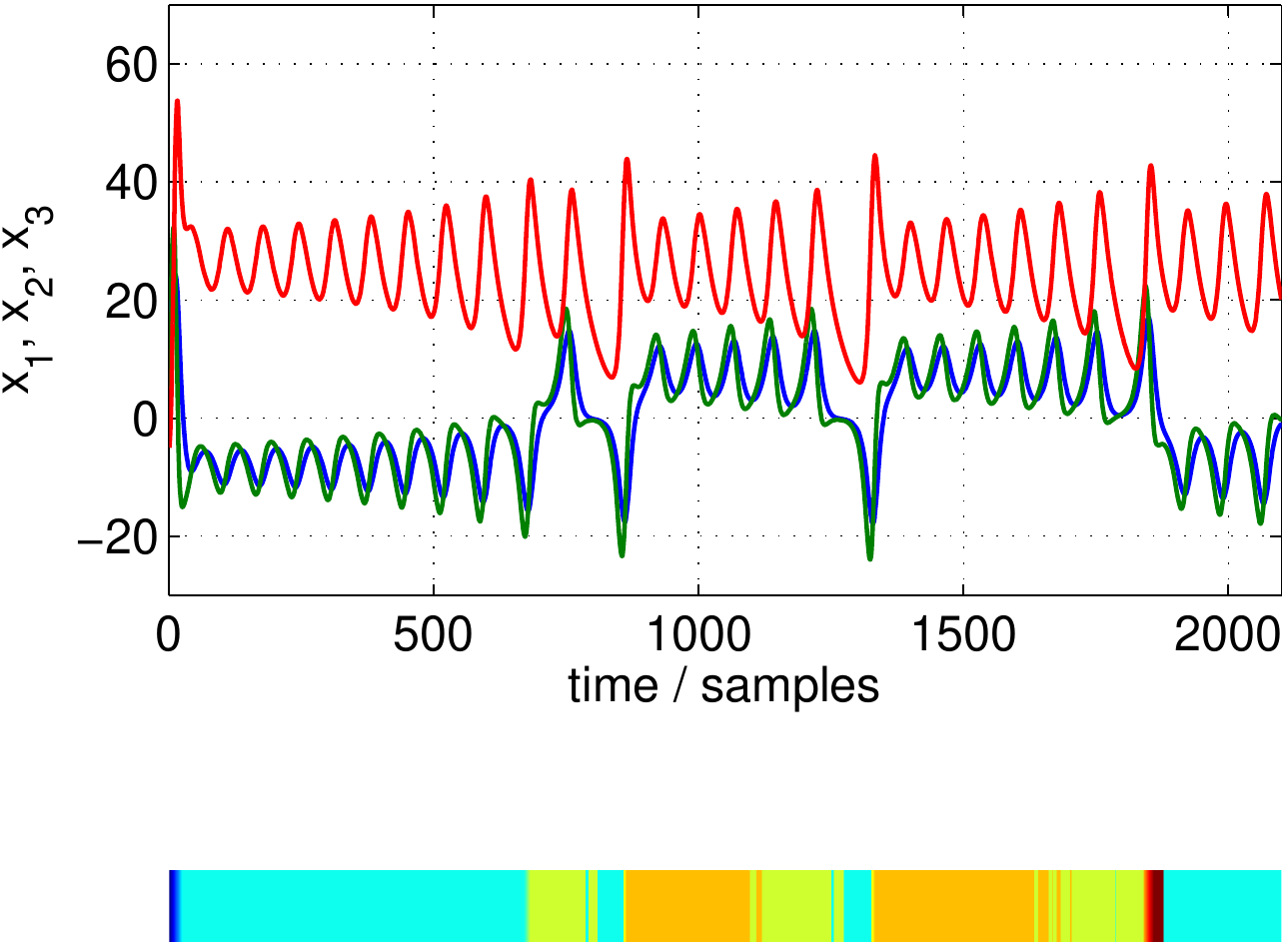}}
\subfigure[]{\includegraphics[width=60mm]{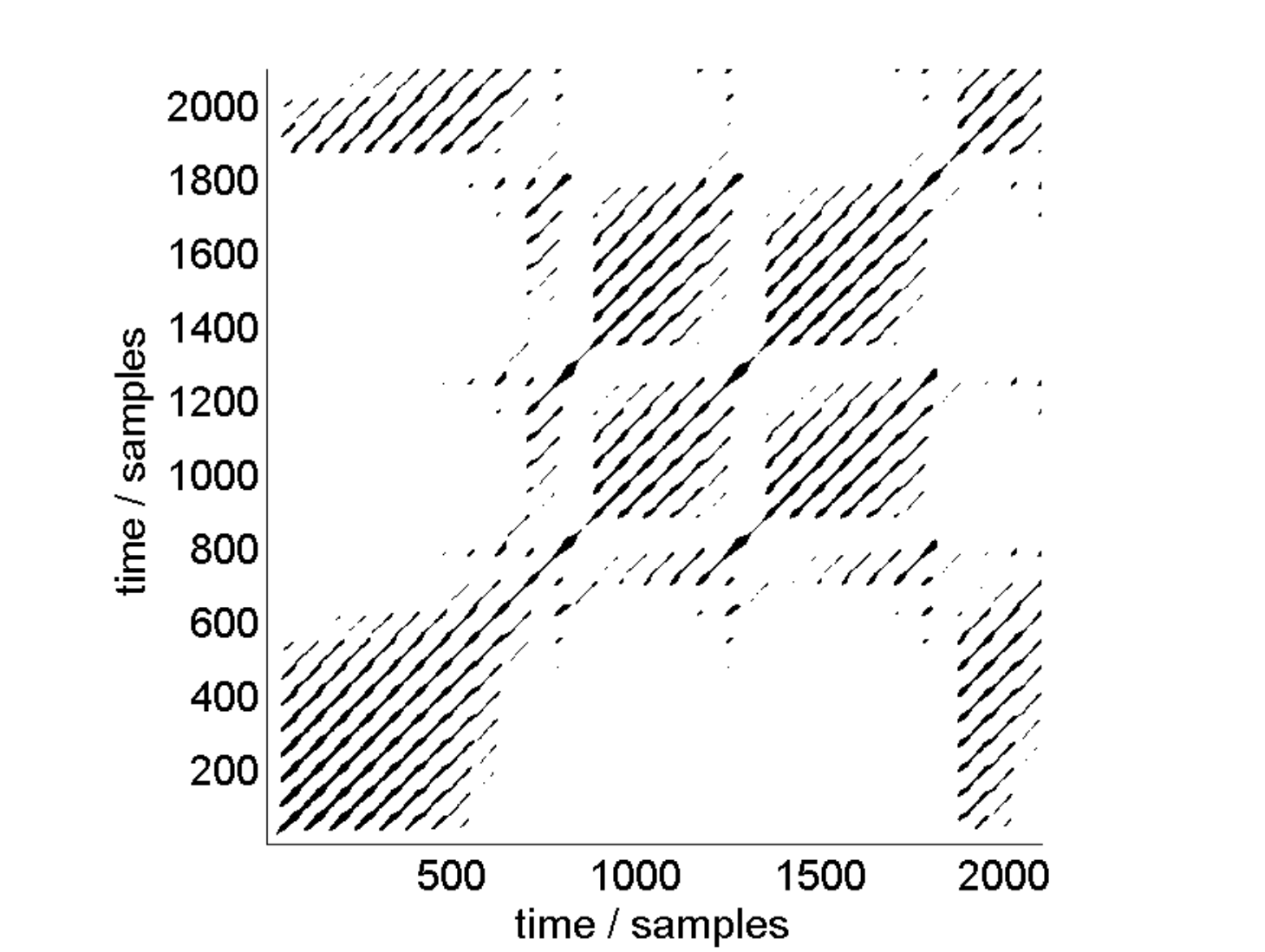}}
\subfigure[]{\includegraphics[width=60mm]{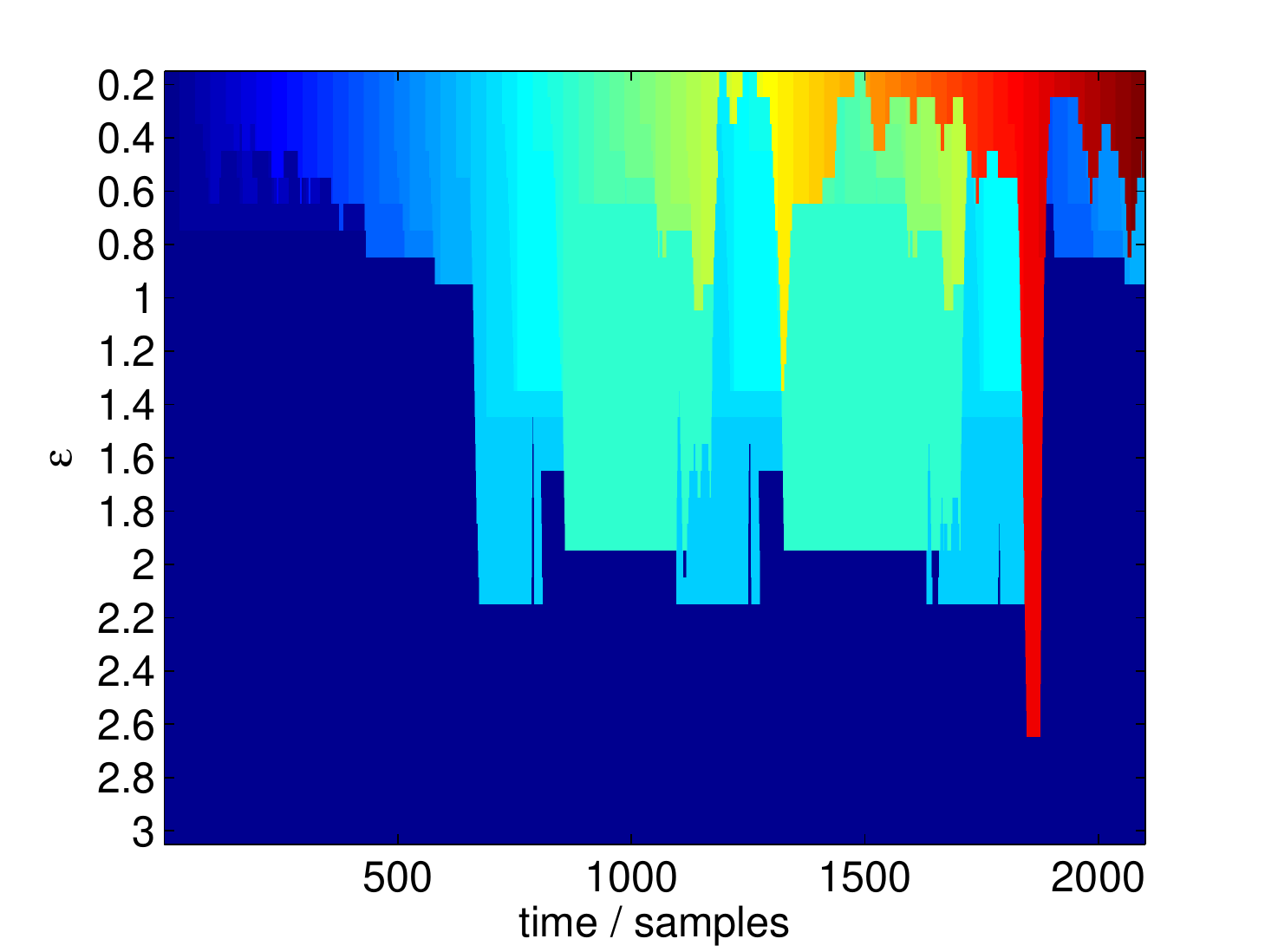}}
\subfigure[]{\includegraphics[width=60mm]{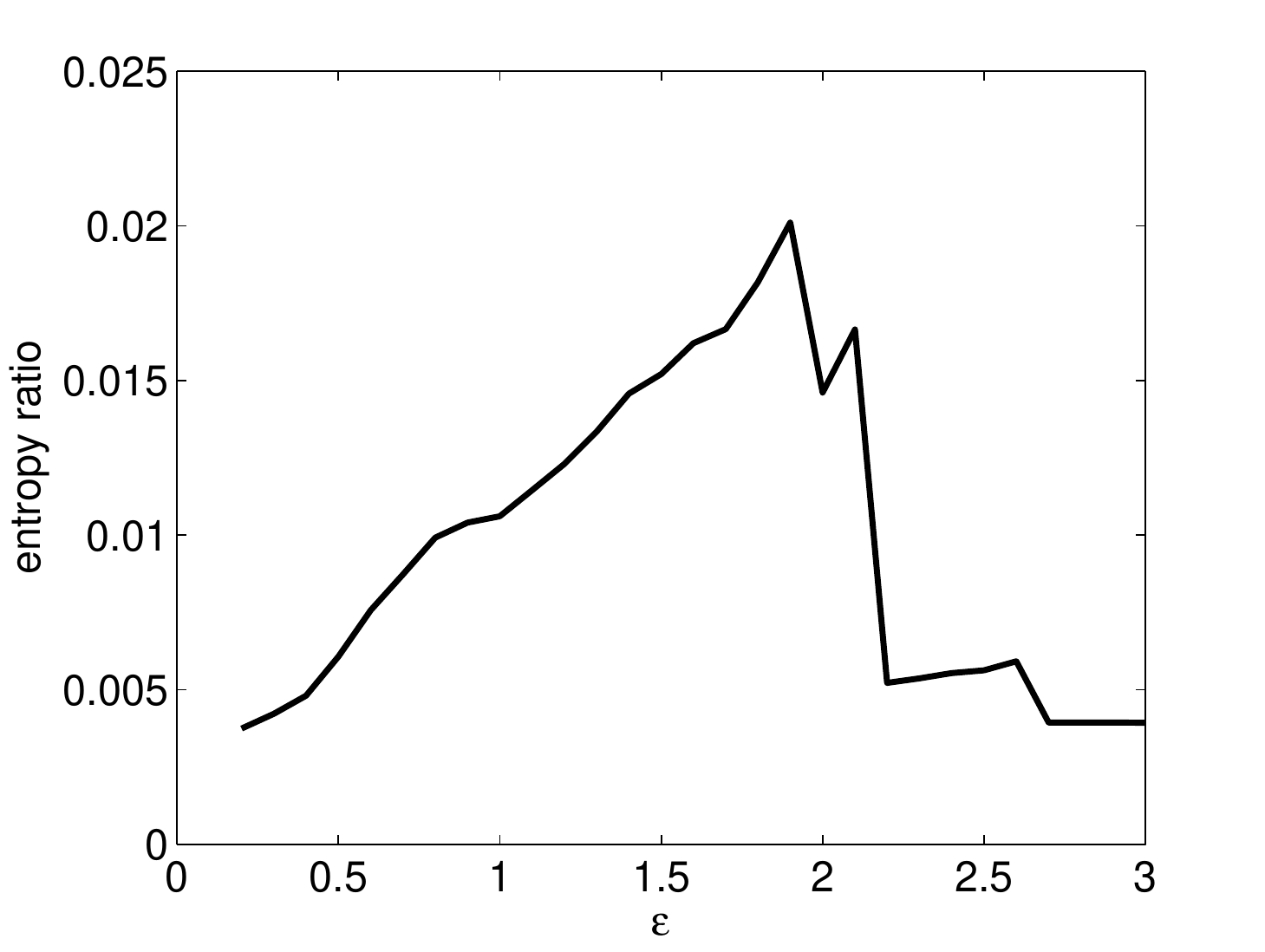}}
\subfigure[]{\includegraphics[width=60mm]{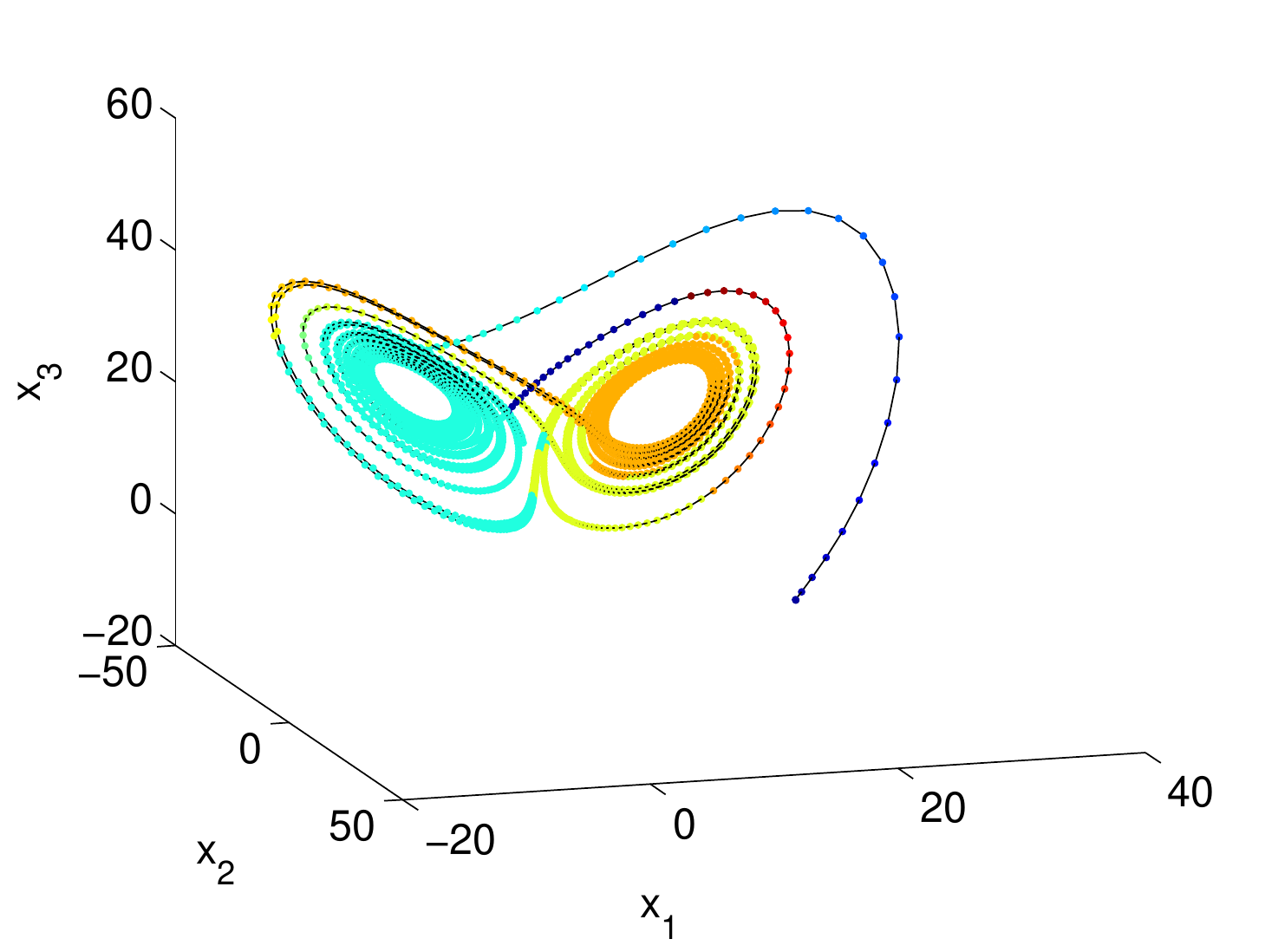}}
\subfigure[]{\includegraphics[width=60mm]{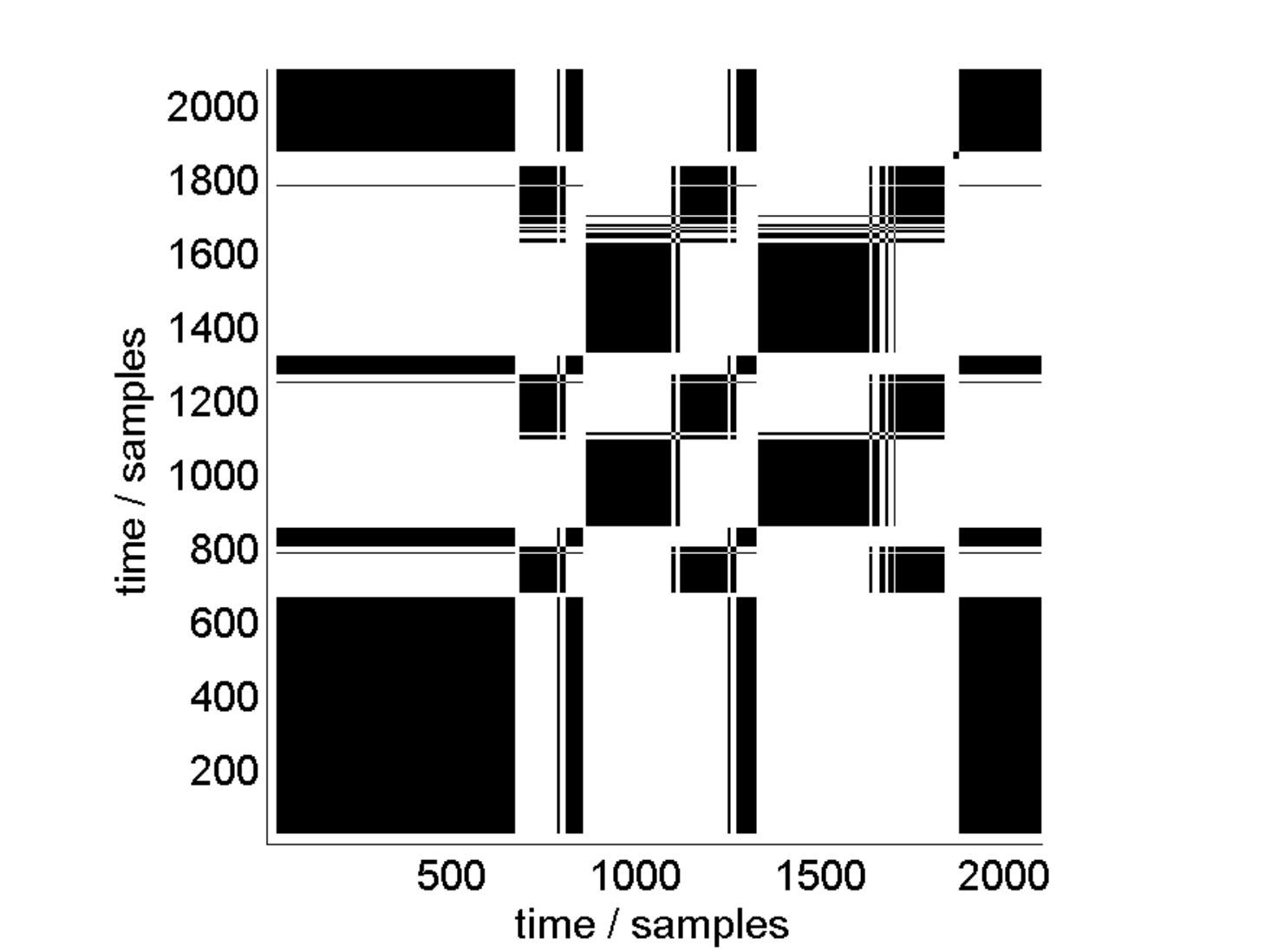}}
\caption{\label{fig:lorenz} (Color online) Recurrence-based symbolic dynamics of the Lorenz attractor \cite{Lorenz63}. (a) Time series $\vec{x}_t$ (upper panel) and optimal encoding $s'$ (color bar beneath). (b) $\varepsilon$-Recurrence plot [\Eq{eq:rp}] for $\varepsilon = 5.0$ and Euclidian norm; black pixels denote $R_{ij} = 1$, white ones $R_{ij} = 0$. (c) Symbolic dynamics $s'$ for range $\varepsilon \in [0.2, 3]$. (d) Dependence of entropy ratio [\Eq{eq:entrat}] from $\varepsilon$. (e) Phase space partition into recurrence domains for optimal encoding $\varepsilon^* = 1.9$. (f) Symbolic recurrence plot [\Eq{eq:rpsym}] of optimal encoding.}
\end{figure}

When $x_j \in \mathbb{R}^d$ is the system's state at (discretized) time $j$ in phase space $\mathbb{R}^d$ of dimension $d$, the element
\begin{equation}\label{eq:rp}
    R_{ij} = \Theta(\varepsilon -  || x_j - x_i ||)
\end{equation}
of the \emph{recurrence matrix} $\vec{R} = (R_{ij})$ is one if $x_j$ is contained in a ``ball'' $B_\varepsilon(x_i)$ of radius $\varepsilon > 0$ centered at state $x_i \in \mathbb{R}^d$ and zero otherwise \cite{EckmannOliffsonRuelle87, MarwanKurths05}, as mediated by the Heaviside step function $\Theta$. Eckmann et al. \cite{EckmannOliffsonRuelle87} have already pointed out that RPs display recurrence domains as a characteristic ``checkerboard texture''. We illustrate this in \Fig{fig:lorenz} with the paradigmatic Lorenz attractor \cite{Lorenz63}.

The upper panel of \Fig{fig:lorenz}(a) displays the $x_1, x_2$ (blue and green) and $x_3$ (red) time series of the Lorenz attractor starting with initial condition $\vec{x} = [20, 5, -5]^T$, with integration interval $[0, 20]$ and sampling $\Delta t = 0.0095$. The two wings of the attractor [shown in \Fig{fig:lorenz}(e)] clearly correspond to positive, respectively, negative $x_1, x_2$. The recurrence plot in \Fig{fig:lorenz}(b) exhibits the typical texture of diagonal line patterns that are characteristic for oscillatory dynamics. These oscillators correspond to the attractor's wings. Going along the line of identity (LOI) reveals transient transitions between about four recurrence domains, while the checkerboard texture of these diagonal line patters indicates that there are indeed only two recurrence domains being involved, namely the wings, that are repeatedly explored by the system's trajectory.

For uniform $\varepsilon$, the recurrence matrices obtained from \Eq{eq:rp} are reflexive, $R_{ii} = 1$ (the LOI), and symmetric, $R_{ij} = R_{ji}$, but in general not transitive, i.e. $R_{ij} = 1$ and $R_{jk} = 1$ do not necessarily imply $R_{ik} = 1$. In order to cope with this disadvantage, Donner et al. and later Faure and Lesne \cite{DonnerHinrichsScholz08, FaureLesne10} suggested to compute the recurrence matrix from words in a symbolic dynamics \cite{Hao89} through
\begin{equation}\label{eq:rpsym}
    R^+_{ij} = \delta_{w_i w_j} \:,
\end{equation}
where $w_i, w_j$ are words of length $m$ at times $i$ and $j$ in a symbolic sequence $s = a_1 a_2 \dots a_n$. Here, $\delta_{ab} = 1$ if $a = b$ and zero otherwise, denotes the Kronecker matrix. Symbolic RPs given by \Eq{eq:rpsym} are also transitive, because symbolic dynamics results from a partition of the system's phase space into equivalence classes from an equivalence relation.

In contrast to \cite{DonnerHinrichsScholz08, FaureLesne10} we here construct a phase space partition and thereby its resulting symbolic dynamics from the $\varepsilon$-RP \pref{eq:rp}. For that aim we first observe that $R_{ij} = 1$ if two $\varepsilon$-balls $B_\varepsilon(x_i)$ and $B_\varepsilon(x_j)$ intersect: $B_\varepsilon(x_i) \cap B_\varepsilon(x_j) \ne \emptyset$. We could therefore start with an initial partition of the phase space into a family of $\varepsilon$-balls around the sampling points $x_i$ and its set complement and then merge all intersecting balls together. The result is a partition of phase space into disjoint sets.

In order to achieve this construction we consider the $\varepsilon$-RP $\vec{R}$ [\Eq{eq:rp}] as a \emph{grammatical rewriting system} over the time indices of a given trajectory $x_t$ \cite{HopcroftUllman79}. Thus, we first map a trajectory $x_t$ to the sequence of successive time indices, regarded as symbols: $x_t \to s_t = t$. Then we define a formal grammar of rewriting rules: if $i > j$ and $R_{ij} = 1$ create a rule $i \to j$. To enforce transitivity for  $i > j > k$, $R_{ij} = 1$ and $R_{ik} = 1$, we first eliminate the redundancy by rewriting only $i \to k$ and then create an additional rule $j \to k$. Finally, we apply this grammar to the initial sequence of time indices $s_t = t$ in order to replace large indices by smaller ones, thus exploiting the recurrence structure of the data. The result is a transformed symbolic sequence $s'_t$, whose symbolic RP $\vec{R}^+$ [\Eq{eq:rpsym}] \cite{DonnerHinrichsScholz08, FaureLesne10} becomes also transitive.

Let us illustrate the procedure by means of a simple example. Assume, we have a series of only five data points $(x_1, x_2, \dots, x_5)$ that gave rise to the recurrence matrix
\begin{equation}\label{eq:examp}
    \vec{R} = \begin{bmatrix}
                1 & 0 & 0 & 1 & 0 \\
                0 & 1 & 0 & 1 & 1 \\
                0 & 0 & 1 & 0 & 0 \\
                1 & 1 & 0 & 1 & 0 \\
                0 & 1 & 0 & 0 & 1 \\
              \end{bmatrix}
\end{equation}
The algorithm starts in the 5th row, detecting a recurrence $R_{52} = 1$. Since $5>2$, we create a rewriting rule $5 \to 2$. Because the next recurrence in row 5 is trivial, the algorithm continues with row 4, where $R_{41} = R_{42} = 1$. Now, two rules $4 \to 1$ and $4 \to 2$ could be generated. However, the latter is redundant. Therefore the algorithm only records the rule $4 \to 1$. Moreover, transitivity is taken into account by an additional rule $2 \to 1$. Next, row 3 does not contribute to the algorithm and rows 2 and 1 can be neglected due to the symmetry. Recursively applying this grammar to the symbolically encoded time series $s = \text{\flqq}12345\text{\frqq}$ yields $s' = \text{\flqq}11311\text{\frqq}$, i.e. a system with two recurrence domains \flqq1\frqq{} and \flqq3\frqq.

In order to validate our construction, we employ the method to the Lorenz attractor as shown in \Fig{fig:lorenz}(c). Here, each row is the symbolically encoded time series $s'$ from \Fig{fig:lorenz}(a) using a color code. For small values of $\varepsilon$ (top rows) there are almost no intersecting $\varepsilon$-balls such that each ball is represented by a separate color from the light spectrum. Increasing $\varepsilon$ towards the bottom rows yields more and more intersections, eventually leading to one big cluster of merged $\varepsilon$-balls for $\varepsilon > 2.6$. For intermediate values of $\varepsilon$ essentially two recurrence domains emerge that are connected by transients.

Interestingly, \Fig{fig:lorenz}(c) also reveals that our recurrence-based symbolic dynamics is rather robust against variations of the ball size $\varepsilon$ which is reflected by the vertical band structure of the symbolic sequences.

Guided by the principle of maximal entropy, we assume that the system spends equal portions of time in its recurrence domains and derive a utility function of the symbolic encoding from the entropy of the symbol distribution
\begin{equation}\label{eq:entropy}
    H(\varepsilon) = - \sum_k^{M(\varepsilon)} p_k \log p_k \:,
\end{equation}
where $p_k$ is the relative frequency of symbol $k$ and $M(\varepsilon)$ the cardinality of the symbolic repertoire obtained for ball size $\varepsilon$. The entropy ratio
\begin{equation}\label{eq:entrat}
    h(\varepsilon) = \frac{H(\varepsilon)}{M(\varepsilon)}
\end{equation}
is then a good estimator for a given encoding because small values of $\varepsilon$ lead to an almost  uniform distribution of rare symbols that is punished by the large alphabet. By contrast, large values of $\varepsilon$ give rise to a trivial partition with small entropy. Thus, the quantity $h(\varepsilon)$ will assume a global maximum for an optimal value
\begin{equation}\label{eq:optcod}
    \varepsilon^* = \arg \max_\varepsilon h(\varepsilon)
\end{equation}
reflecting a uniform distribution of a small number of recurrence domains.

We plot the dependence of $h(\varepsilon)$ for the Lorenz system in \Fig{fig:lorenz}(d) and choose the optimal ball size $\varepsilon^* = 1.9$ for the symbolic dynamics in the color bar of \Fig{fig:lorenz}(a). One can easily recognize that one wing is uniquely represented by the turquois symbol, while the other one is represented by orange and light green symbols. The distribution of these symbols in phase space is shown in \Fig{fig:lorenz}(e) using the same color palette for the samples $\vec{x}_t$. Here, one wing is completely captured by the union of turquois $\varepsilon$-balls, while the other one needs two partitions cells, indicated in orange and light green which is due to a gap in the second wing in our numerics.

Finally, \Fig{fig:lorenz}(f) depicts the symbolic RP \Eq{eq:rpsym} where the characteristic checkerboard texture of the Lorenz attractor's recurrence domains is significantly enhanced.

\begin{figure}[H]
\centering
\subfigure[]{\includegraphics[width=60mm]{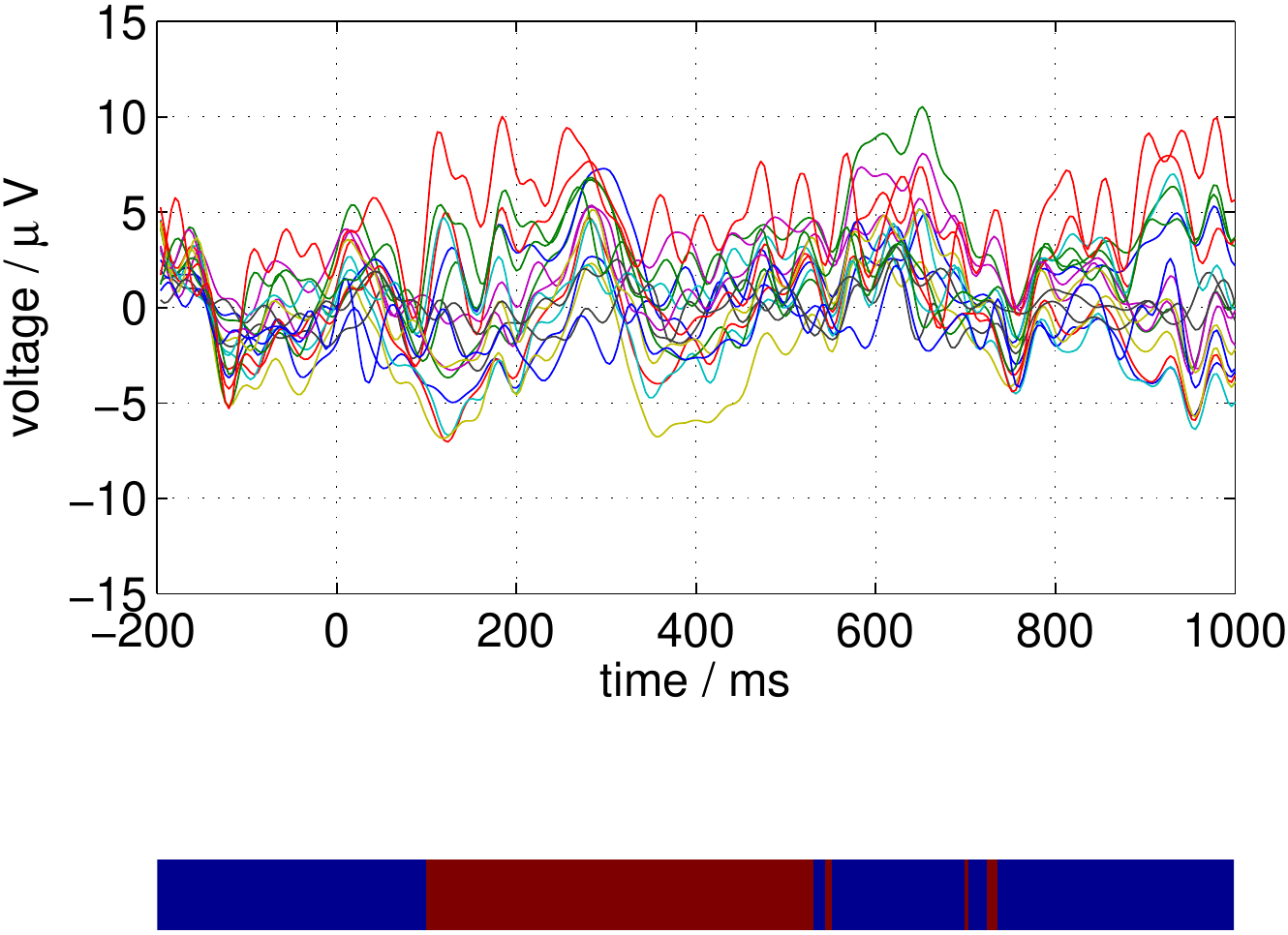}}
\subfigure[]{\includegraphics[width=60mm]{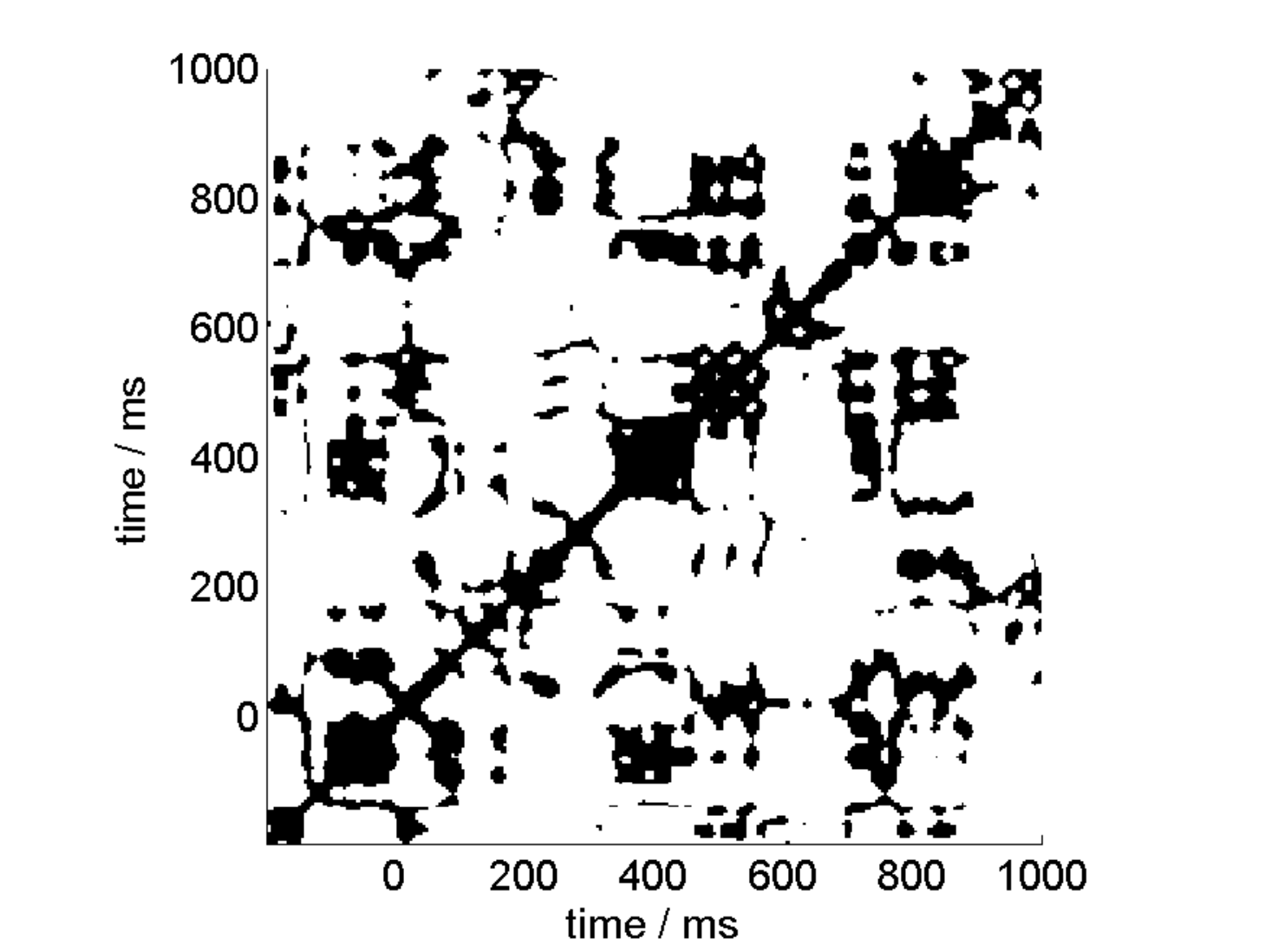}}
\subfigure[]{\includegraphics[width=60mm]{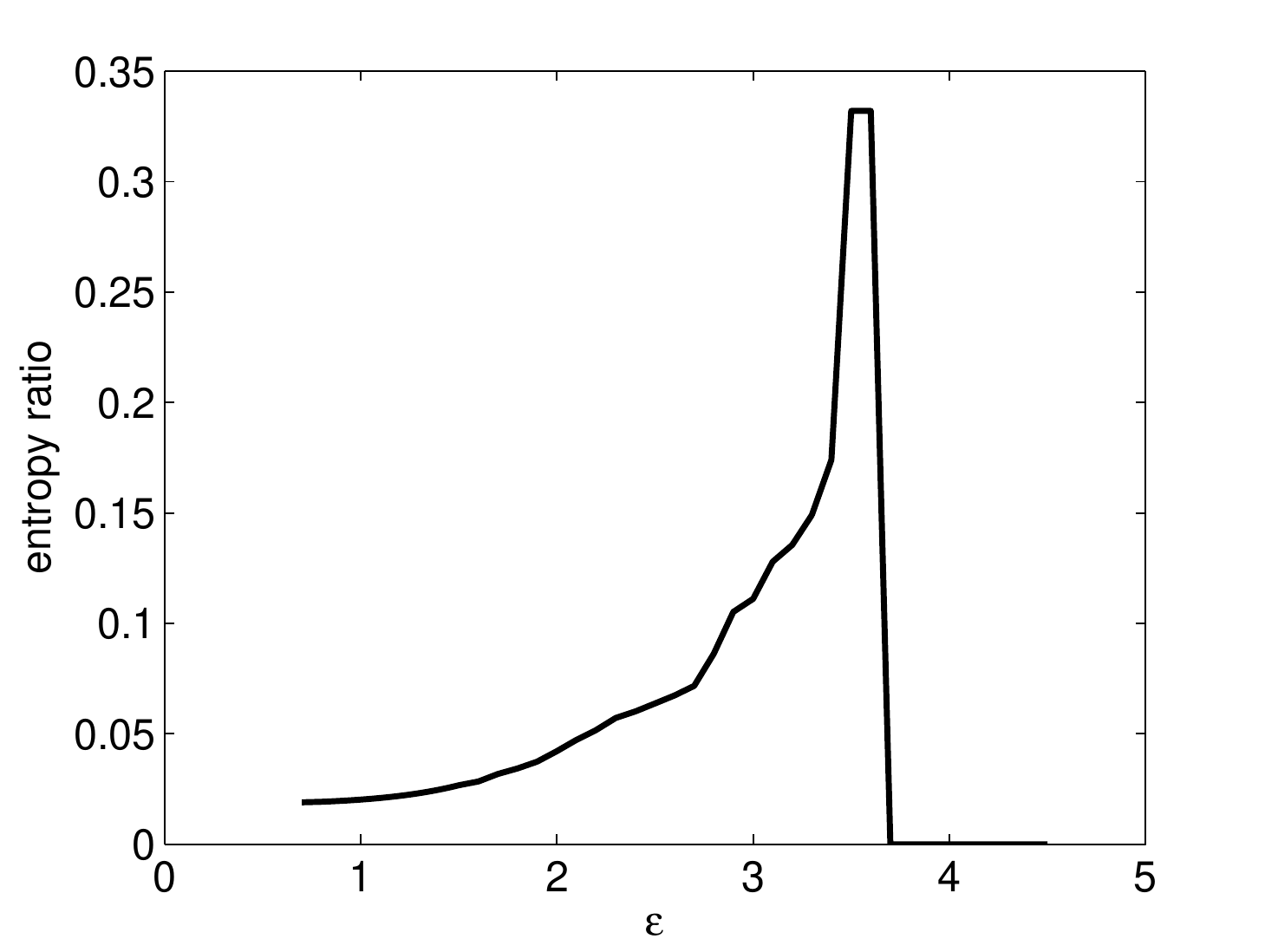}}
\subfigure[]{\includegraphics[width=60mm]{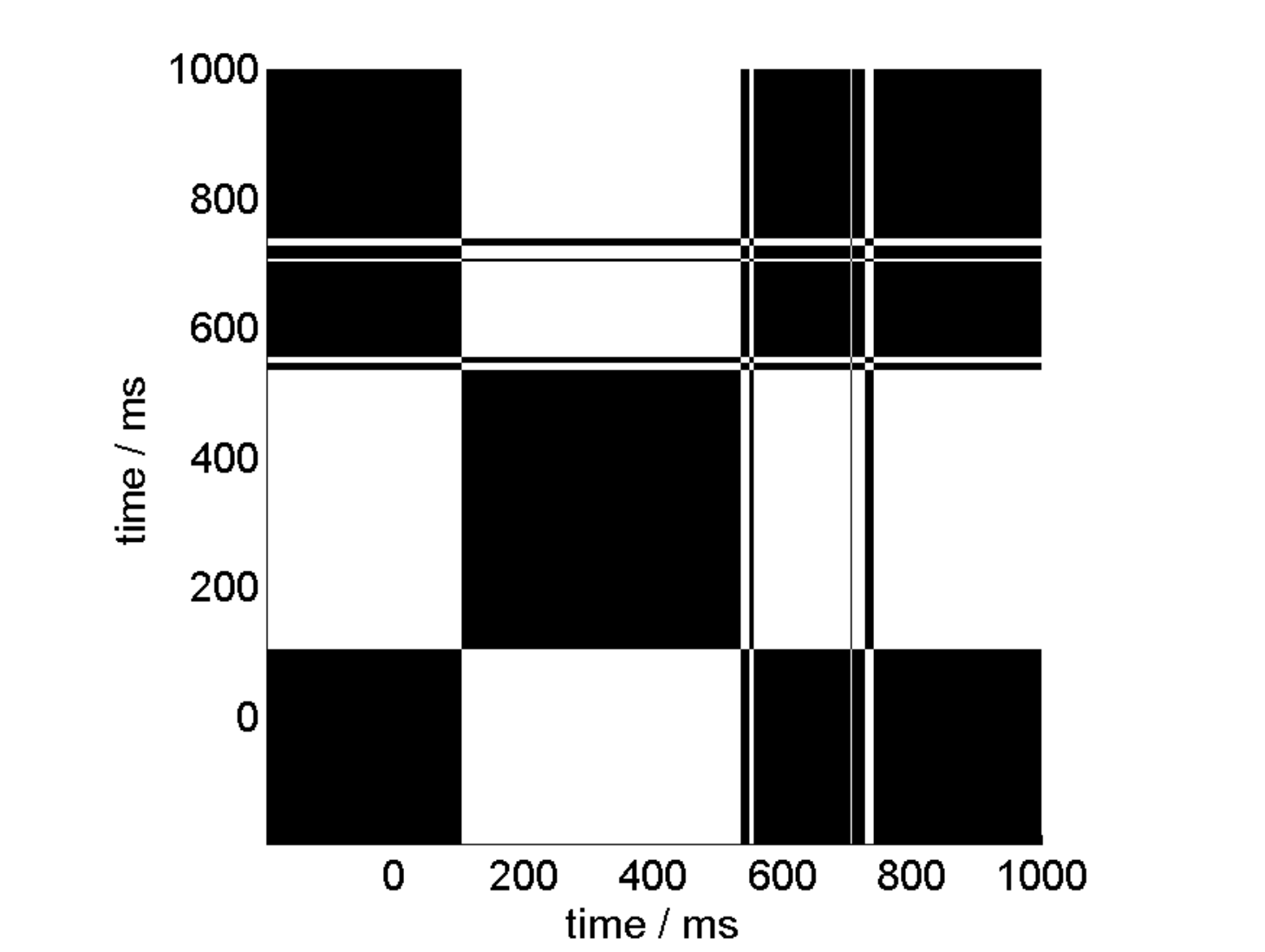}}
\caption{\label{fig:erp} (Color online) Recurrence-based symbolic dynamics of a $17$-dimensional ERP data set of a single subject \cite{GrabenSaddyEA00}. (a) Single subject ERP time series (upper panel) and symbolic encoding $s'$ with $\varepsilon^* = \quant{3.5}{\mu V}$ (color bar beneath) for $d = 17$ scalp channels. (b) $\varepsilon$-Recurrence plot [\Eq{eq:rp}] for $\varepsilon = \quant{8.0}{\mu V}$. (c) Utility function $h(\varepsilon)$ [\Eq{eq:entrat}]. (d) Symbolic recurrence plot [\Eq{eq:rpsym}] for optimal encoding $\varepsilon^* = \quant{3.5}{\mu V}$.}
\end{figure}

In order to also present a proof-of-concept for our method applied to real-world data, we reanalyze event-related electroencephalographic (EEG) data from a language processing experiment \cite{GrabenSaddyEA00} in \Fig{fig:erp} since Hutt and Riedel \cite{HuttRiedel03} have argued that components in the event-related brain potential (ERP) can be regarded as saddle sets and therefore as recurrence domains in the EEG.

Figure \ref{fig:erp}(a) displays the averaged ERP time series of a single subject encountering a linguistic processing problem in German \cite{GrabenSaddyEA00}. In \Fig{fig:erp}(b) we present the conventional RP [\Eq{eq:rp}] for $\varepsilon = \quant{8.0}{\mu V}$. Figure \ref{fig:erp}(c) shows the utility function $h(\varepsilon)$ from \Eq{eq:entrat}. The optimal encoding $s'$ is obtained for $\varepsilon^* = \quant{3.5}{\mu V}$ which is depicted as the color bar in \Fig{fig:erp}(a). The symbolic dynamics exhibits three interesting properties: \emph{i}) the pre-stimulus interval is represented by one distinguished recurrence domain; \emph{ii}) the time interval for lexical access around 400 ms post stimulus is represented by another recurrence domain; \emph{iii}) the time window of syntactic reanalysis around 600 ms is represented by the first recurrence domain again. This results from the optimization constraint to obtain uniformly distributed recurrence domains. Also the symbolic RP in \ref{fig:erp}(d) nicely reveals the existence of two substantial recurrence domains.

In this Letter we proposed a parsimonious algorithm for the detection of recurrence domains in complex dynamical systems. In contrast to techniques based on Markov chains, which require an \emph{ad hoc} partitioning of the system's phase space and the estimation of transition probabilities, our approach exploits the recurrence structure of the system's dynamics thereby partitioning the phase space into unions of intersecting $\varepsilon$-balls along the actual trajectory.

The proposed method could have a number of interesting applications in many different fields, such as molecular dynamics \cite{DeuflhardWeber05}, geo- \cite{FroylandPadbergEA07}, and neurosciences \cite{AllefeldAtmanspacherWackermann09, Friston97a, RabinovichHuertaLaurent08, RabinovichEA08} for the identification of recurrence domains. Moreover, it could also be useful for the analysis of complex networks for solving graph partition and related problems by taking the transitive closure of the graph's adjacency matrix. Finally we concede that further research is required to obtain appropriate utility functions for real-world problems that violate the uniformity assumption for recurrence domains in order to detect e.g. saddle sets in heteroclinic dynamics \cite{RabinovichEA08, RabinovichHuertaLaurent08} or to identify functional ERP components \cite{HuttRiedel03}.

This research has been supported by the European Union's Seventh Framework Programme (FP7/2007-2013) ERC grant agreement No. 257253 awarded to AH, hosting PbG during fall 2012 in Nancy, and by a Heisenberg fellowship (GR 3711/1-1) of the German Research Foundation (DFG) awarded to PbG.


\begin{thebibliography}{20}
\expandafter\ifx\csname natexlab\endcsname\relax\def\natexlab#1{#1}\fi
\expandafter\ifx\csname bibnamefont\endcsname\relax
  \def\bibnamefont#1{#1}\fi
\expandafter\ifx\csname bibfnamefont\endcsname\relax
  \def\bibfnamefont#1{#1}\fi
\expandafter\ifx\csname citenamefont\endcsname\relax
  \def\citenamefont#1{#1}\fi
\expandafter\ifx\csname url\endcsname\relax
  \def\url#1{\texttt{#1}}\fi
\expandafter\ifx\csname urlprefix\endcsname\relax\def\urlprefix{URL }\fi
\providecommand{\bibinfo}[2]{#2}
\providecommand{\eprint}[2][]{\url{#2}}

\bibitem[{\citenamefont{Larralde and Leyvraz}(2005)}]{LarraldeLeyvraz05}
\bibinfo{author}{\bibfnamefont{H.}~\bibnamefont{Larralde}} \bibnamefont{and}
  \bibinfo{author}{\bibfnamefont{F.}~\bibnamefont{Leyvraz}},
  \bibinfo{journal}{Physical Review Letters} \textbf{\bibinfo{volume}{94}},
  \bibinfo{eid}{160201} (\bibinfo{year}{2005}).

\bibitem[{\citenamefont{Hutt and Riedel}(2003)}]{HuttRiedel03}
\bibinfo{author}{\bibfnamefont{A.}~\bibnamefont{Hutt}} \bibnamefont{and}
  \bibinfo{author}{\bibfnamefont{H.}~\bibnamefont{Riedel}},
  \bibinfo{journal}{Physica D} \textbf{\bibinfo{volume}{177}},
  \bibinfo{pages}{203 } (\bibinfo{year}{2003}).

\bibitem[{\citenamefont{Rabinovich
  et~al.}(2008{\natexlab{a}})\citenamefont{Rabinovich, Huerta, Varona, and
  Afraimovich}}]{RabinovichEA08}
\bibinfo{author}{\bibfnamefont{M.~I.} \bibnamefont{Rabinovich}},
  \bibinfo{author}{\bibfnamefont{R.}~\bibnamefont{Huerta}},
  \bibinfo{author}{\bibfnamefont{P.}~\bibnamefont{Varona}}, \bibnamefont{and}
  \bibinfo{author}{\bibfnamefont{V.~S.} \bibnamefont{Afraimovich}},
  \bibinfo{journal}{PLoS Computational Biology} \textbf{\bibinfo{volume}{4}},
  \bibinfo{pages}{e1000072} (\bibinfo{year}{2008}{\natexlab{a}}).

\bibitem[{\citenamefont{Lorenz}(1963)}]{Lorenz63}
\bibinfo{author}{\bibfnamefont{E.~N.} \bibnamefont{Lorenz}},
  \bibinfo{journal}{Journal of the Atmospheric Sciences}
  \textbf{\bibinfo{volume}{20}}, \bibinfo{pages}{130 } (\bibinfo{year}{1963}).

\bibitem[{\citenamefont{Poincar\'e}(1890)}]{Poincare1890}
\bibinfo{author}{\bibfnamefont{H.}~\bibnamefont{Poincar\'e}},
  \bibinfo{journal}{Acta Mathematica} \textbf{\bibinfo{volume}{13}},
  \bibinfo{pages}{1 } (\bibinfo{year}{1890}).

\bibitem[{\citenamefont{Deuflhard and Weber}(2005)}]{DeuflhardWeber05}
\bibinfo{author}{\bibfnamefont{P.}~\bibnamefont{Deuflhard}} \bibnamefont{and}
  \bibinfo{author}{\bibfnamefont{M.}~\bibnamefont{Weber}},
  \bibinfo{journal}{Linear Algebra and its Applications}
  \textbf{\bibinfo{volume}{398}}, \bibinfo{pages}{161 } (\bibinfo{year}{2005}).

\bibitem[{\citenamefont{Froyland et~al.}(2007)\citenamefont{Froyland, Padberg,
  England, and Treguier}}]{FroylandPadbergEA07}
\bibinfo{author}{\bibfnamefont{G.}~\bibnamefont{Froyland}},
  \bibinfo{author}{\bibfnamefont{K.}~\bibnamefont{Padberg}},
  \bibinfo{author}{\bibfnamefont{M.~H.} \bibnamefont{England}},
  \bibnamefont{and} \bibinfo{author}{\bibfnamefont{A.~M.}
  \bibnamefont{Treguier}}, \bibinfo{journal}{Physical Review Letters}
  \textbf{\bibinfo{volume}{98}}, \bibinfo{pages}{224503}
  (\bibinfo{year}{2007}).

\bibitem[{\citenamefont{Allefeld et~al.}(2009)\citenamefont{Allefeld,
  Atmanspacher, and Wackermann}}]{AllefeldAtmanspacherWackermann09}
\bibinfo{author}{\bibfnamefont{C.}~\bibnamefont{Allefeld}},
  \bibinfo{author}{\bibfnamefont{H.}~\bibnamefont{Atmanspacher}},
  \bibnamefont{and}
  \bibinfo{author}{\bibfnamefont{J.}~\bibnamefont{Wackermann}},
  \bibinfo{journal}{Chaos} \textbf{\bibinfo{volume}{19}},
  \bibinfo{pages}{015102} (\bibinfo{year}{2009}).

\bibitem[{\citenamefont{Friston}(1997)}]{Friston97a}
\bibinfo{author}{\bibfnamefont{K.~J.} \bibnamefont{Friston}},
  \bibinfo{journal}{NeuroImage} \textbf{\bibinfo{volume}{5}},
  \bibinfo{pages}{164 } (\bibinfo{year}{1997}).

\bibitem[{\citenamefont{Rabinovich
  et~al.}(2008{\natexlab{b}})\citenamefont{Rabinovich, Huerta, and
  Laurent}}]{RabinovichHuertaLaurent08}
\bibinfo{author}{\bibfnamefont{M.~I.} \bibnamefont{Rabinovich}},
  \bibinfo{author}{\bibfnamefont{R.}~\bibnamefont{Huerta}}, \bibnamefont{and}
  \bibinfo{author}{\bibfnamefont{G.}~\bibnamefont{Laurent}},
  \bibinfo{journal}{Science} \textbf{\bibinfo{volume}{321}}, \bibinfo{pages}{48
  } (\bibinfo{year}{2008}{\natexlab{b}}).

\bibitem[{\citenamefont{Froyland}(2005)}]{Froyland05}
\bibinfo{author}{\bibfnamefont{G.}~\bibnamefont{Froyland}},
  \bibinfo{journal}{Physica D} \textbf{\bibinfo{volume}{200}}, \bibinfo{pages}{205}
  (\bibinfo{year}{2005}).

\bibitem[{\citenamefont{Gaveau and Schulman}(2006)}]{GaveauSchulman06}
\bibinfo{author}{\bibfnamefont{B.}~\bibnamefont{Gaveau}} \bibnamefont{and}
  \bibinfo{author}{\bibfnamefont{L.~S.} \bibnamefont{Schulman}},
  \bibinfo{journal}{Physical Reviews E} \textbf{\bibinfo{volume}{73}},
  \bibinfo{eid}{036124} (\bibinfo{year}{2006}).

\bibitem[{\citenamefont{Eckmann et~al.}(1987)\citenamefont{Eckmann, Kamphorst,
  and Ruelle}}]{EckmannOliffsonRuelle87}
\bibinfo{author}{\bibfnamefont{J.-P.} \bibnamefont{Eckmann}},
  \bibinfo{author}{\bibfnamefont{S.~O.} \bibnamefont{Kamphorst}},
  \bibnamefont{and} \bibinfo{author}{\bibfnamefont{D.}~\bibnamefont{Ruelle}},
  \bibinfo{journal}{Europhysics Letters} \textbf{\bibinfo{volume}{4}},
  \bibinfo{pages}{973 } (\bibinfo{year}{1987}).

\bibitem[{\citenamefont{Marwan and Kurths}(2005)}]{MarwanKurths05}
\bibinfo{author}{\bibfnamefont{N.}~\bibnamefont{Marwan}} \bibnamefont{and}
  \bibinfo{author}{\bibfnamefont{J.}~\bibnamefont{Kurths}},
  \bibinfo{journal}{Physics Letters} \textbf{\bibinfo{volume}{A 336}},
  \bibinfo{pages}{349 } (\bibinfo{year}{2005}).

\bibitem[{\citenamefont{Donner et~al.}(2008)\citenamefont{Donner, Hinrichs, and
  Scholz-Reiter}}]{DonnerHinrichsScholz08}
\bibinfo{author}{\bibfnamefont{R.}~\bibnamefont{Donner}},
  \bibinfo{author}{\bibfnamefont{U.}~\bibnamefont{Hinrichs}}, \bibnamefont{and}
  \bibinfo{author}{\bibfnamefont{B.}~\bibnamefont{Scholz-Reiter}},
  \bibinfo{journal}{The European Physical Journal Special Topics}
  \textbf{\bibinfo{volume}{164}}, \bibinfo{pages}{85 } (\bibinfo{year}{2008}).

\bibitem[{\citenamefont{Faure and Lesne}(2010)}]{FaureLesne10}
\bibinfo{author}{\bibfnamefont{P.}~\bibnamefont{Faure}} \bibnamefont{and}
  \bibinfo{author}{\bibfnamefont{A.}~\bibnamefont{Lesne}},
  \bibinfo{journal}{International Journal of Bifurcation and Chaos}
  \textbf{\bibinfo{volume}{20}}, \bibinfo{pages}{1731 } (\bibinfo{year}{2010}).

\bibitem[{\citenamefont{Hao}(1989)}]{Hao89}
\bibinfo{author}{\bibfnamefont{B.-L.} \bibnamefont{Hao}},
  \emph{\bibinfo{title}{Elementary Symbolic Dynamics and Chaos in Dissipative
  Systems}} (\bibinfo{publisher}{World Scientific},
  \bibinfo{address}{Singapore}, \bibinfo{year}{1989}).

\bibitem[{\citenamefont{Hopcroft and Ullman}(1979)}]{HopcroftUllman79}
\bibinfo{author}{\bibfnamefont{J.~E.} \bibnamefont{Hopcroft}} \bibnamefont{and}
  \bibinfo{author}{\bibfnamefont{J.~D.} \bibnamefont{Ullman}},
  \emph{\bibinfo{title}{Introduction to Automata Theory, Languages, and
  Computation}} (\bibinfo{publisher}{Addison--Wesley}, \bibinfo{address}{Menlo
  Park, California}, \bibinfo{year}{1979}).

\bibitem[{\citenamefont{beim Graben et~al.}(2000)\citenamefont{beim Graben,
  Saddy, Schlesewsky, and Kurths}}]{GrabenSaddyEA00}
\bibinfo{author}{\bibfnamefont{P.}~\bibnamefont{beim Graben}},
  \bibinfo{author}{\bibfnamefont{D.}~\bibnamefont{Saddy}},
  \bibinfo{author}{\bibfnamefont{M.}~\bibnamefont{Schlesewsky}},
  \bibnamefont{and} \bibinfo{author}{\bibfnamefont{J.}~\bibnamefont{Kurths}},
  \bibinfo{journal}{Physical Reviews E} \textbf{\bibinfo{volume}{62}},
  \bibinfo{pages}{5518 } (\bibinfo{year}{2000}).

\end{thebibliography}
\end{document}